\documentclass[aps,pre,floatfix,groupedaddress,amsmath,amssymb,twocolumn]{revtex4-2}
\usepackage{graphicx}
\usepackage{dcolumn}
\usepackage{bm}
\usepackage{hyperref}
\usepackage{booktabs}
\usepackage{siunitx}
\usepackage{url}
\newcommand{\chidf}[2]{\makebox[2em][r]{#1}/\makebox[1em][l]{#2}}

\begin{document}

\title{Fractals of Simple Random Walks in Two Dimensions: A Monte Carlo Study}

\author{Jiang Zhou}
\thanks{These two authors contributed equally to this paper.}
\affiliation{Department of Physics, Guizhou University, Guiyang 550025, China}
\author{Ziru Deng}
\thanks{These two authors contributed equally to this paper.}
\affiliation{Department of Information Science and Technology, Donghua University, Shanghai 200051, China}
\author{Pengcheng Hou}
\email{houpc@hfnl.cn}
\affiliation{Hefei National Laboratory, Hefei 230088, China}

\date{\today}

\begin{abstract}
    We present a Monte Carlo study of the fractal geometry of clusters formed by discrete-time simple random walks (sRW) of $L^2$ steps on a periodic square $L\times L$ lattice.
    We verify with high precision that the asymptotic behavior of the cluster mass follows $M/L^2 \simeq  (\ln L)^{-1} [\frac{\pi}{2}+b (\ln L)^{-2}]$, with $b\approx -(\pi/2)^{-2}$, demonstrating marginal “logarithmic fractals”.
    We further determine the fractal dimension of the hull to be $d_{\rm hull}=1.333\,29(14)=4/3$, in excellent agreement with the prediction of Schramm-Loewner evolution ($\rm SLE_{8/3}$) for the Brownian frontier universality class.
    More importantly, we analyze the chemical distance $S$ spanning the cluster and obtain strong evidence that it asymptotically scales as $S\sim L(\ln L)^{1/4}$, lying exactly on the theoretical upper bound for the chemical distance for level-set percolation clusters on the two-dimensional Gaussian free field. Our numerical results show that the sRW cluster exhibits a conformally invariant external frontier and contains highly efficient asymptotically linear connective paths.
\end{abstract}

\maketitle

\section{Introduction}
Random walk is one of the most fundamental stochastic processes in science and engineering\cite{spitzer1976,lawlerlimic2010,hughes1995,revesz2013}. Even in its simplest lattice form, the model captures the essential mechanisms of Brownian motion, diffusion, and transport in disordered media\cite{montrollweiss1965,havlin1987,benavraham2000}.
Random walk describes fluctuation noise in electronic systems, search strategies in biology, polymer conformations\cite{degennes1979}, and it underlies a broad class of algorithmic sampling procedures. In physics, random walk is also a rich geometric object in its own right---the trace, frontier, and hole structures of a random walk trajectory are naturally described in fractal language\cite{mandelbrot1982,deng2010}. Because of this breadth, the random walk serves as a universal testing ground for geometric and statistical ideas that are difficult to isolate in more complicated interacting systems.

The random walk is equally central in mathematics and field theory, where many of its basic properties change qualitatively with spatial dimension\cite{lawlerlimic2010}. Recurrence, intersection probabilities, cover times, range statistics, and scaling limits behave very differently in one, two, and higher dimensions\cite{erdostaylor1960,jainpruitt1970,schramm2000,lawler2008}. These dimension-dependent features govern not only dynamical observables but also the geometry of the trace and its boundary. The two-dimensional (2D) case is especially delicate because it is marginal: the walk is recurrent, yet its trace forms a ramified, weakly space-filling cluster punctured by holes on many length scales. As a result, logarithmic corrections play an essential role, and several geometric questions remain subtle, notably how to classify the resulting fractal structure and how intrinsic shortest paths scale with system size\cite{soares2025}.

Two geometric observables are particularly informative. The first is the hull perimeter, whose scaling $\mathcal{L}\sim L^{d_{\rm hull}}$ probes the external frontier and thus the conformal universality class of the cluster boundary\cite{lawler2001,werner2004}. The second is the chemical distance, which measures the shortest path inside the cluster and probes the efficiency of internal connectivity\cite{havlin1985,zhou2012}. These two observables are complementary: the hull is sensitive to the outer geometry, while the chemical distance detects how tortuous the internal network is. Together they provide a sharper characterization than the cluster mass alone.

This perspective is relevant to recent work on the chemical distance in level-set percolation on the two-dimensional Gaussian free field (GFF)\cite{schramm2009,ding2018,ding2020,ding2022}. For crossings between distant boundaries, Ding, Wirth, and collaborators proved that the chemical distance is at most of order $L(\ln L)^{1/4}$\cite{ding2020}. Whether this upper bound is sharp, and whether additional $(\ln\ln L)^m$ factors are genuinely present, are open questions. Since the trace of the 2D sRW is another paradigmatic sparse random geometry with nontrivial structure---and one connected to the GFF through isomorphism theorems relating random walk local times to the GFF \cite{dynkin1983,sznitman2012,lejan2011,lupu2016}---it is natural to ask whether its shortest-path scaling exhibits the same nearly linear behavior.

In this work we focus on two questions. First, how should the geometric cluster formed by a 2D sRW of $L^2$ steps be classified from its mass and hull scaling? Second, what is the asymptotic scaling behavior of the spanning chemical distance across that cluster? To answer these questions, we perform large-scale Monte Carlo simulations on periodic square lattices and analyze the finite-size scaling of the cluster mass $M$, the hull perimeter $\mathcal{L}$, and a spanning chemical distance $S$. For the mass, we do not merely confirm the leading Dvoretzky-Erd\H{o}s behavior; we also resolve the finite-size correction structure under periodic boundary conditions. Our data yield
\begin{equation}
    M\simeq\frac{L^2}{\ln L}\left[\frac{\pi}{2}-(\frac{\pi}{2}\ln L)^{-2}\right].
\end{equation}
The leading term $\pi L^2/(2\ln L)$ confirms the expected marginally fractal scaling. More importantly, whereas the infinite-plane analysis suggests a subleading $(\ln L)^{-1}$ correction structure\cite{csaki2024}, our PBC data do not resolve such a contribution and are instead dominated by a visible correction of order $(\ln L)^{-2}$. For the hull, we obtain $\mathcal{L}\sim L^{d_{\rm hull}}$ with $d_{\rm hull}=4/3$, in agreement with the $\mathrm{SLE}_{8/3}$ prediction for the Brownian frontier\cite{schramm2000,lawler2001,lawler2001b,lawler2001c,lawler2002,werner2004,beffara2008,cardy2005}; the leading finite-size correction is well described by a term $\sim 1/L$. For the spanning chemical distance, our data support $S\sim L(\ln L)^{1/4}$ over the accessible size up to $L=2^{16}$, without evidence for an additional $(\ln\ln L)^m$ factor. These three results together show that the 2D sRW cluster is marginally fractal in mass, has a Brownian-frontier-type exterior, and contains asymptotically efficient internal paths.

The remainder of the paper is organized as follows. Section II defines the model and the observables. Section III presents the finite-size scaling analysis and the numerical results. Section IV discusses the implications of the results and summarizes the main conclusions.

\begin{figure}[t]
    \centering
    \includegraphics[width=0.8\linewidth]{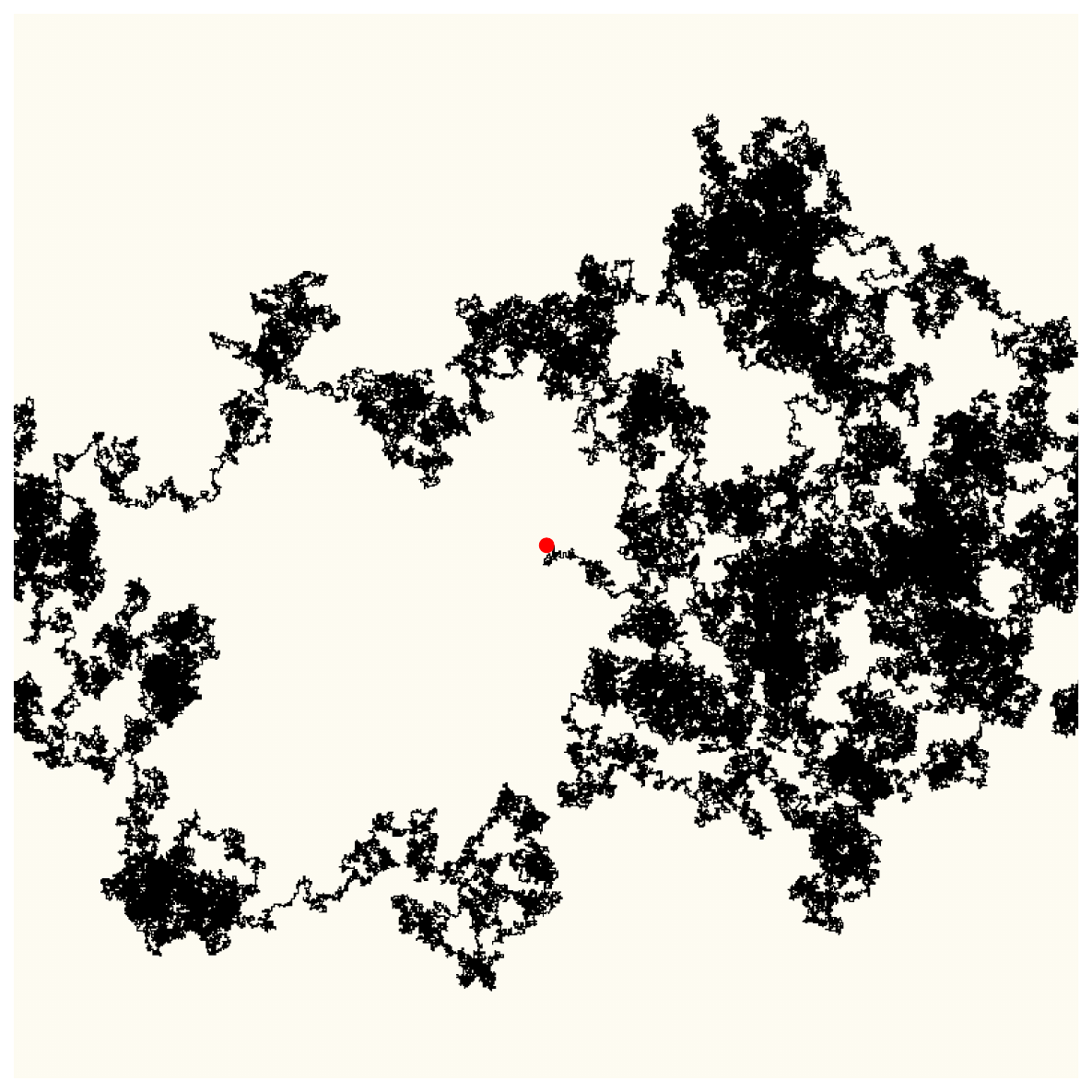}
    \caption{\label{fig1} Typical sRW trajectory of $L^2$ steps on a periodic $L\times L$ square lattice with $L=800$, started from the red point $(0,0)$. The trace is weakly space filling and leaves holes on all length scales.}
\end{figure}

\section{Model and Observables}
We study a discrete-time sRW on a two-dimensional periodic square lattice of linear size $L$. The walker starts from a prescribed seed site and, at each time step, moves with equal probability to one of the four nearest neighbors. PBCs are imposed in both directions, so the walk takes place on a discrete torus of volume $L^2$.

Throughout the paper, the walk time (or walk steps) is fixed to $N=L^2$. This is well below the cover time, which scales as $\simeq (1/\pi)L^2(\ln L)^2$ on the two-dimensional torus\cite{dembo2004}, and probes the marginal regime where the trajectory has explored the entire system scale but still leaves a nonvanishing hierarchy of unvisited regions. Figure~\ref{fig1} illustrates a typical hierarchical structure on a square lattice with $L=800$.
There are two natural definitions of the cluster generated by a random walk. In the bond-cluster definition, the cluster consists of the visited sites together with the bonds actually traversed by the walk, so connectivity is determined by the traced path itself. In the site-cluster definition, the cluster is defined only by the set of visited sites, and connectivity is assigned afterward through nearest-neighbor adjacency. This distinction matters for observables that depend explicitly on graph connectivity. In particular, quantities such as the chemical distance and cluster-number statistics can depend on which definition is adopted, whereas observables determined only by the visited-site set, such as the mass and site visitation probabilities, are independent of this choice.

In the following we use the bond-cluster definition throughout. With this convention, both the hull perimeter and the spanning chemical distance are defined directly on the traced bond cluster, while the mass remains unchanged. Since our interest is in the asymptotic scaling behavior, the conclusions drawn below are not tied to microscopic differences between the bond- and site-cluster definitions. For each independent trajectory, we measure the following three observables.

\textit{Cluster mass $M$.} The mass $M$ is the number of distinct lattice sites visited by the walk up to time $N=L^2$. In two dimensions, recurrence leads to many revisits\cite{csaki2024}, so $M$ grows more slowly than the total number of steps and provides the most direct measure of the weak noncompactness of the trace.

\textit{Hull perimeter $\mathcal{L}$.} The hull is the interface separating the bond cluster from the background ``ocean''. The algorithm proceeds as follows. First, on the dual lattice, we identify all connected components of the complement of the cluster. The ``ocean'' is defined as the largest such connected component, which on the torus corresponds to the dominant unvisited region; all remaining complement components are classified as internal holes. Next, coastline dual vertices are identified as those ocean vertices that share at least one dual edge with a non-ocean region. Finally, the hull perimeter $\mathcal{L}$ is obtained by tracing the boundary along these coastline vertices and counting the dual edges that lie at the interface between the ocean and the filled cluster. Simply put, the hull is the outer boundary (or frontier) of the geometric cluster\cite{grassberger1986,grossman1986,voss1984}.

\textit{Spanning chemical distance $S$.} The chemical distance is the length of the shortest path constrained to lie on the cluster\cite{havlin1985,zhou2012}. Here we use a spanning version: starting from the seed site $\mathbf{r}_0$, we perform a breadth-first search on the bond cluster and define $S$ as the maximal graph distance reached within the connected trace. This quantity measures the scale of the most efficient paths needed to span the cluster.

Unless stated otherwise, the symbols $M$, $\mathcal{L}$, and $S$ denote ensemble averages over independent sRW realizations. The largest systems studied have $L=2^{16}=65536$ and the walk time $N=2^{32}$ steps.

\section{Numerical Results}

The numerical analysis is based on independent sRW trajectories on periodic square lattices with linear sizes up to $L=2^{16}$. For each size, we generate trajectories of length $N=L^2$ and average the observables defined in Sec. II.
The statistics decrease with system size from $3.3\times 10^8$ samples for the small lattices to $4\times 10^5$ samples at $L=65536$. All error bars reported below are standard errors of the mean computed from independent realizations. The total samples for each $L$ are divided into independent batches, and the batch-to-batch variance determines the statistical uncertainty. Nonlinear least-squares fits are weighted by these errors, and parameter uncertainties are obtained from the covariance matrix at the minimum. We focus below on the finite-size scaling behaviors extracted from these data.

\begin{figure}[t]
\centering
\includegraphics[width=0.45\textwidth]{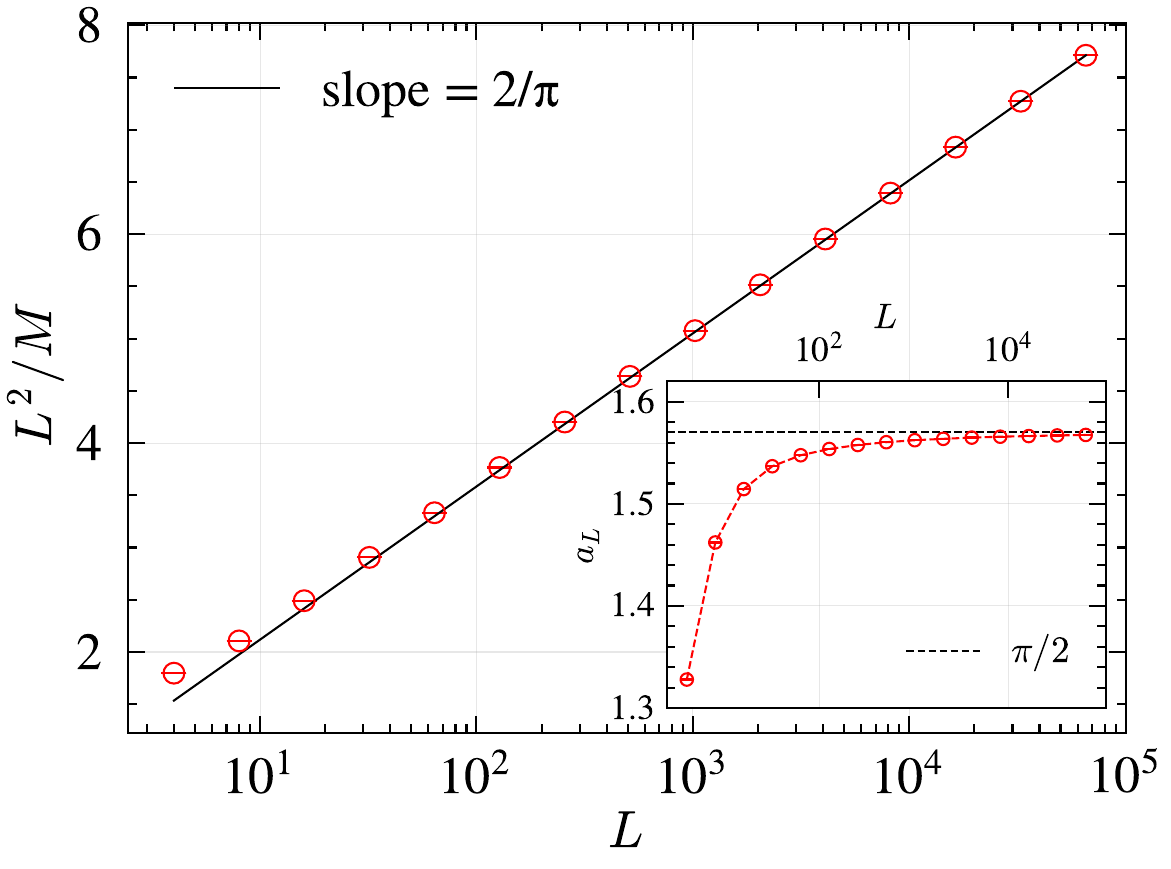}
\caption{\label{mass1} Rescaled cluster mass $L^2/M$ versus $\ln L$. The large-$L$ data approach a straight line of slope $2/\pi$, consistent with $M\sim (\pi/2)L^2/\ln L$. The inset shows the approach of $a_L$ to $\pi/2$.}
\end {figure}

\subsection{Cluster Mass \texorpdfstring{$M$}{M}}
For a 2D sRW of $N=L^2$ steps, the Dvoretzky-Erd\H{o}s theorem gives the leading asymptotic behavior $M\sim (\pi/2)L^2/\ln L$ in the infinite-plane limit\cite{dvoretzky1951}. Our Monte Carlo data are fully consistent with this prediction: the semi-logarithmic plot of $L^2/M$ versus $\ln L$ approaches a straight line with slope $2/\pi$, as shown in Fig.~\ref{mass1}.
In order to quantify finite-size effects under PBCs, we write
\begin{equation}
    a_L=\frac{M\ln L}{L^2},
\end{equation}
which tends to $\pi/2$ in the large-$L$ limit, as shown in the inset of Fig.~\ref{mass1}. We then fit the data to the finite-size ansatz
\begin{equation}
    M(L)= \frac{L^2}{c+\ln L}[a+b{(\ln L)^{-w}}+\cdots],
    \label{massscale}
\end{equation}
where $a$ and $b$ are amplitudes, $c$ is a logarithmic shift, and $w$ is the subleading correction exponent. The fit quality is assessed by the statistic value $\chi^2/\mathrm{DF}$, where $\mathrm{DF}$ denotes the number of degrees of freedom\cite{hou2019}.
A fit is acceptable if $\chi^2/\mathrm{DF}\simeq 1$.
The results in Table \ref{table_mass} show that the best asymptotic description is obtained for $w=2$. The fitted amplitude converges to $a=1.5714(5)$, in excellent agreement with $\pi/2$, while the shift parameter tends to $c\simeq 1$. This supports a marginally fractal geometry: the mass has effective dimension $2$, but the space-filling tendency is suppressed by logarithmic corrections. Such kinds of geometric objects have been named  “logarithmic fractals”\cite{cao2021,kundu2013,csaki2024}. 

\begin{table}[b]
    \caption{\label{table_mass}Fits of the cluster mass $M$ to Eq.~\eqref{massscale}. The consistency of the estimates and the collapse in Fig.~\ref{mass2} support the choice $w=2$.}
    \begin{ruledtabular}
        \begin{tabular}{c c S[table-format=1.4(1)] S[table-format=1.4(1)] S[table-format=-1.4(2)] c}
            {$L_{\min}$} & {$\chi^2/\mathrm{DF}$} & {$c$}     & {$a$}     & {$b$}       & {$w$} \\
            \midrule
            32           & \chidf{10.6}{9}        & 0.997(3)  & 1.5705(3) & -0.412(7)   & 2     \\
            64           & \chidf{2.7}{8}         & 1.004(4)  & 1.5712(4) & -0.386(12)  & 2     \\
            128          & \chidf{2.4}{7}         & 1.006(5)  & 1.5714(5) & -0.378(20)  & 2     \\
            \midrule
            32           & \chidf{12.0}{10}       & 0.9998(4) & {$\pi/2$} & -0.405(3)   & 2     \\
            64           & \chidf{3.8}{9}         & 1.0006(5) & {$\pi/2$} & -0.397(4)   & 2     \\
            128          & \chidf{3.7}{8}         & 1.0005(6) & {$\pi/2$} & -0.399(5)   & 2     \\
            \midrule
            32           & \chidf{12.3}{11}       & {1}       & {$\pi/2$} & -0.4034(9)  & 2     \\
            64           & \chidf{5.7}{10}        & {1}       & {$\pi/2$} & -0.4020(11) & 2     \\
            128          & \chidf{4.3}{9}         & {1}       & {$\pi/2$} & -0.4027(12) & 2     \\
        \end{tabular}
    \end{ruledtabular}
\end{table}

\begin{figure}[t]
    \centering
    \includegraphics[width=0.476\textwidth]{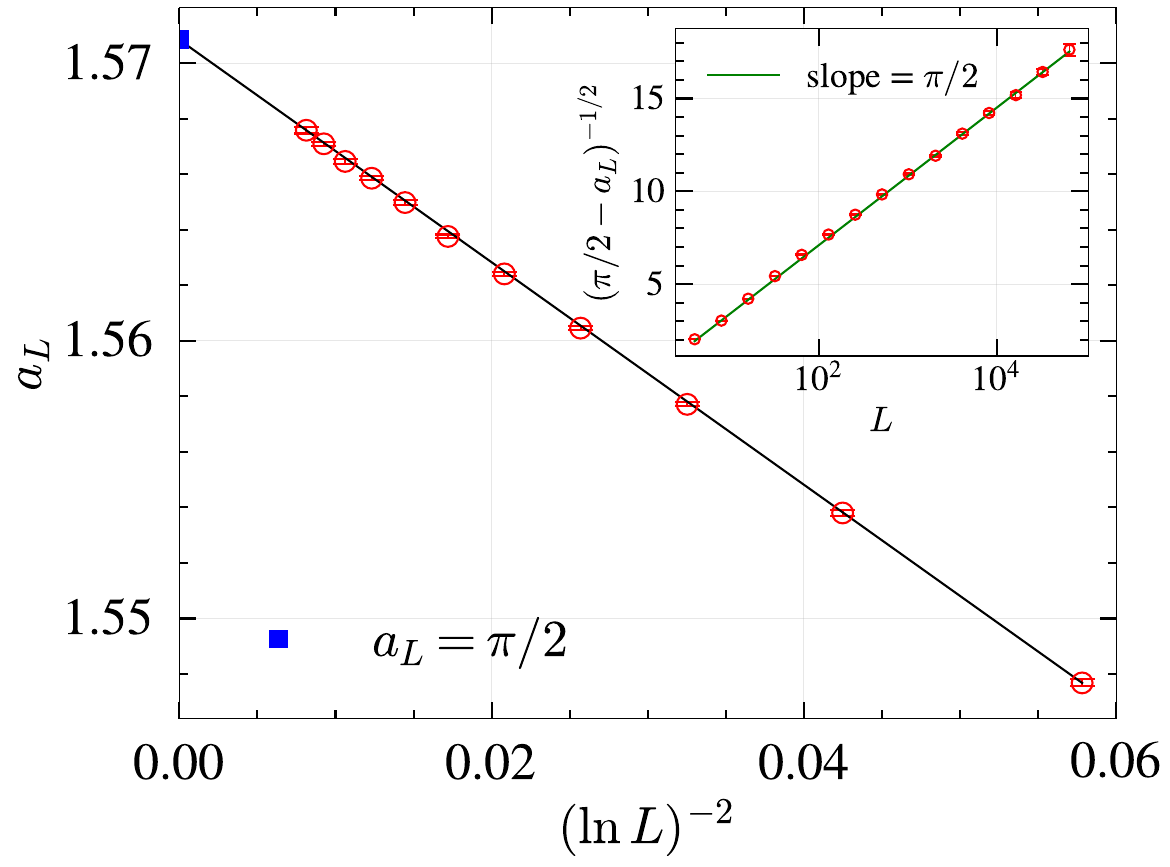}
    \caption{\label{mass2} Collapse of $a_L$ versus $(\ln L)^{-2}$. The asymptotic linearity provides direct graphical support for $w=2$. Inset: test of $(\pi/2-a_L)^{-1/2}\propto \ln L$, consistent with the fitted correction amplitude.}
\end{figure}

The fitting results in Table~\ref{table_mass} also suggest that the leading finite-size correction term is $\simeq b/\ln^2 L$, instead of $\propto 1/\ln L$. Furthermore, we observe that the fitted value of the amplitude $b$ is consistent with $-(2/\pi)^2$.
A more direct test is obtained from the scaling of the already-defined quantity $a_L$. From Eq.~\eqref{massscale}, one expects
\begin{equation}
    a_L = a+b{(\ln L)^{-w}}+\cdots.
    \label{al}
\end{equation}
Figure~\ref{mass2} shows that the data collapse cleanly onto a straight line when $w=2$ is used. This is the main empirical reason for adopting the $(\ln L)^{-2}$ correction. The inset provides an additional cross-check because Eq.~\eqref{al} implies $\pi/2-a_L\simeq (\pi/2)^{-2}(\ln L)^{-2}$, or equivalently $(\pi/2-a_L)^{-1/2}\simeq (\pi/2)\ln L$. The observed linearity is again consistent with the same correction form.

With this evidence, a compact representation of the finite-size behavior is given by
\begin{equation}
    M=\frac{L^2}{1+\ln L}\left[\frac{\pi}{2}-(\frac{\pi}{2}\ln L)^{-2}\right].
\end{equation}
Here the term $(\pi \ln L/2)^{-2}$ should be understood as a finite-size ansatz inferred from the numerical collapse, not as an independent exact theorem. More generally, one expects several logarithmic corrections to coexist in finite systems.
In particular, the infinite-plane result discussed in Ref.~\cite{csaki2024} suggests a subleading $(\ln L)^{-1}$ correction structure. Our PBC data do not resolve such a term within numerical precision; instead, the dominant visible correction is of order $(\ln L)^{-2}$ with amplitude $b=-(\pi/2)^{-2}$.
This suggests that PBCs do not alter the leading mass scaling, but can modify the observable subleading correction structure compared with the infinite-plane case.

\begin{figure}[t]
    \centering
    \includegraphics[width=0.45\textwidth]{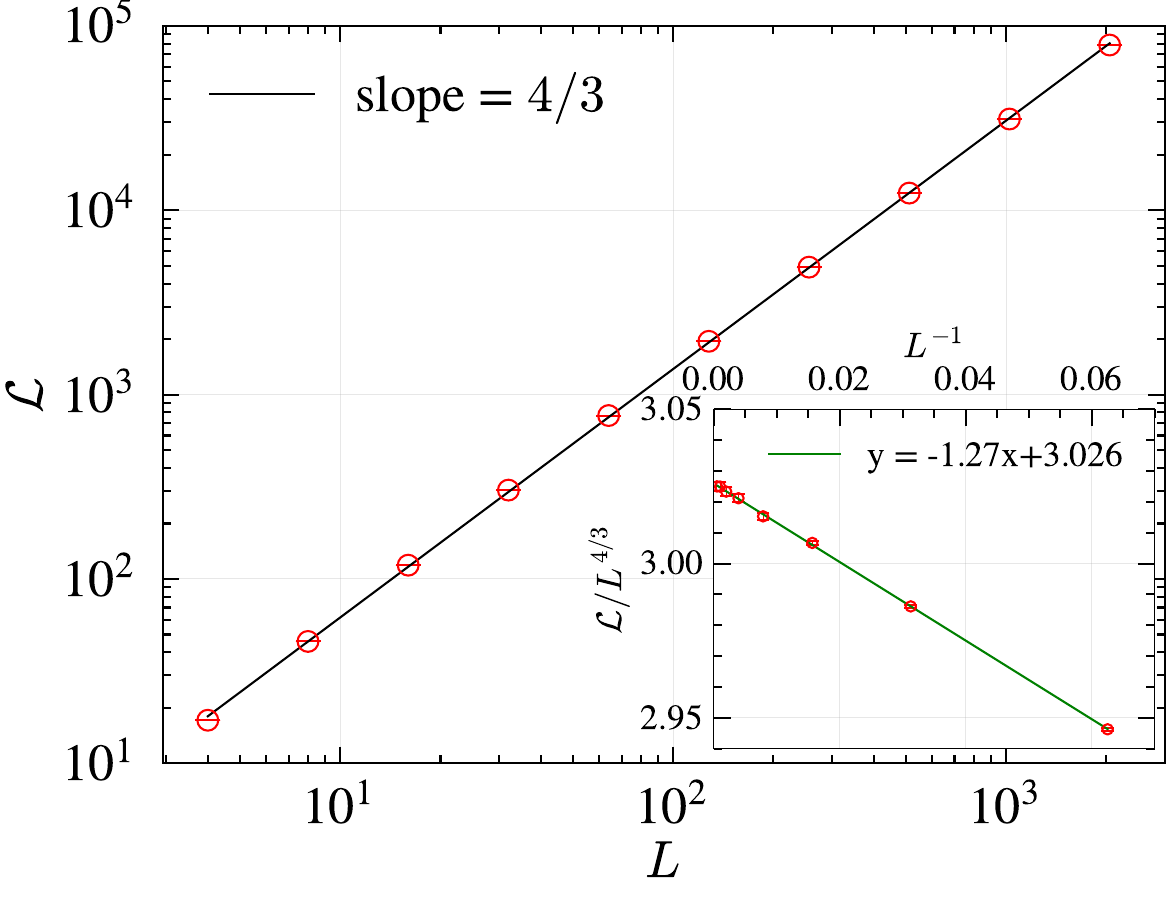}
    \caption{\label{hull}Plot of $\ln \mathcal{L}$ versus $\ln L$. The asymptotic linear behavior is consistent with $\mathcal{L}\sim L^{4/3}$. Inset: collapse of the rescaled hull perimeter $\mathcal{L}/L^{4/3}$ versus $L^{-1}$.}
\end{figure}

\subsection{Hull Perimeter \texorpdfstring{$\mathcal{L}$}{L}}
We next examine the external frontier of the sRW cluster. Figure~\ref{hull} shows that the hull perimeter has a clean power-law dependence on system size,
\begin{equation}
    \mathcal{L}  \sim L^{d_{\rm hull}},
\end{equation}
where $d_{\rm hull}$ is the hull fractal dimension.
We fit the data to
\begin{equation}
    \mathcal{L}(L)=L^{d_{\rm hull}}(a+bL^{-w}),
\end{equation}
with a subleading correction $L^{-w}$. As summarized in Table \ref{hull_fit}, the data are well described by $w=1$, and the estimates stabilize quickly as $L_{\min}$ increases. The optimal estimate is
\begin{align}
    d_{\rm hull}=1.333\,29(14),
\end{align}
in excellent agreement with the exact value $4/3$ for the Brownian frontier\cite{lawler1996,lawler2001}, which is described by the $\mathrm{SLE}_{8/3}$ curve\cite{schramm2000,werner2004}.
Further support comes from the inset of Fig.~\ref{hull}, which shows that the rescaled quantity
$\mathcal{L}/L^{4/3}$ varies linearly with $L^{-1}$, This behavior is the hallmark of a scaling ansatz where the leading exponent is exactly $4/3$ and the dominant correction is $L^{-1}$.

This result is important for two reasons. First, unlike the mass, the hull shows no sign of marginal logarithmic scaling. Second, the estimate $d_{\rm hull}=4/3$ places the outer boundary of the sRW cluster in the Brownian frontier universality class. Any influence of PBCs is therefore confined, within present accuracy, to subleading corrections rather than to the leading exponent.
\begin{table}[htbp]
    \centering
    \caption{Fits for the hull perimeter $\mathcal{L}$. The data support a subleading correction exponent $w=1$.}
    \label{hull_fit}
    \begin{ruledtabular}
        \begin{tabular}{c c S[table-format=1.5(2)] S[table-format=1.4(1)] S[table-format=-1.3(2)] c}
            {$L_{\min}$} & $\chi^2/\mathrm{DF}$ & {$d_{\rm hull}$} & {$a$}      & {$b$}      & {$w$} \\
            \midrule
            4            & \chidf{11.7}{7}      & 1.33302(9)       & 3.0315(11) & -1.327(5)  & 1     \\
            8            & \chidf{1.9}{6}       & 1.33322(11)      & 3.0281(16) & -1.300(10) & 1     \\
            16           & \chidf{1.3}{5}       & 1.33329(14)      & 3.027(3)   & -1.28(3)   & 1     \\
            \midrule
            4            & \chidf{26.6}{8}      & {4/3}            & 3.0275(4)  & -1.314(3)  & 1     \\
            8            & \chidf{3.1}{7}       & {4/3}            & 3.0264(4)  & -1.291(6)  & 1     \\
            16           & \chidf{1.3}{6}       & {4/3}            & 3.0260(5)  & -1.276(13) & 1     \\
            32           & \chidf{1.3}{5}       & {4/3}            & 3.0260(7)  & -1.27(3)   & 1     \\
        \end{tabular}
    \end{ruledtabular}
\end{table}

\begin{figure}[b]
    \centering
    \includegraphics[width=0.93\linewidth]{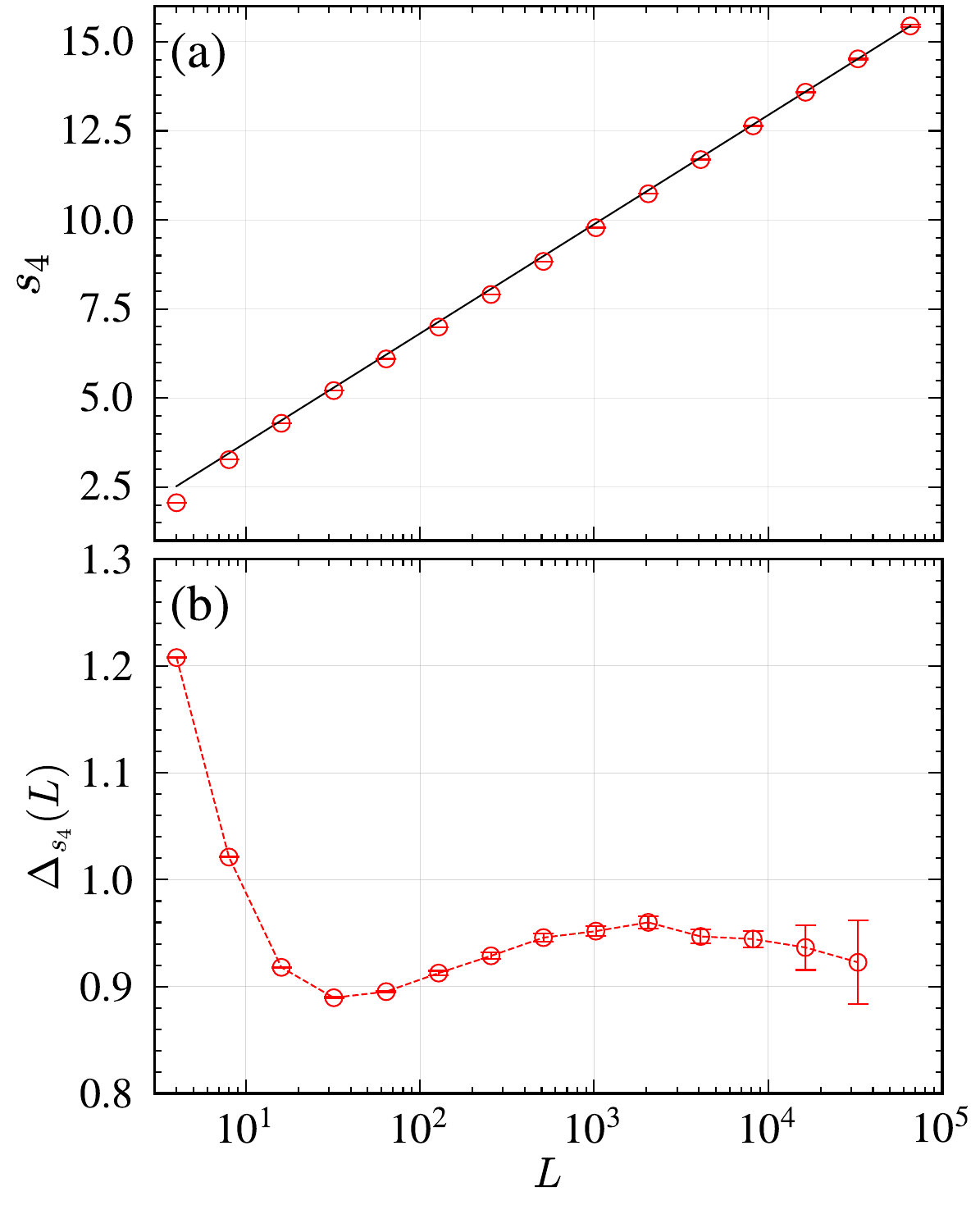}
    \caption{\label{s4_gap} Finite-size scaling of the chemical distance $S$. (a) Collapse of $s_4\equiv (S/L)^4$ versus $\ln L$. (b) Gap function $\Delta_{s_4}(L)$ versus $\ln L$. The gap function shows neither clear signs of asymptotic convergence nor evidence of divergence within the accessible range of system sizes.}
\end{figure}

\subsection{Chemical Distance \texorpdfstring{$S$}{S}}
The chemical distance probes the efficiency of connectivity inside the cluster\cite{havlin1987,benavraham2000}. If the shortest paths were genuinely fractal in the usual sense, one would expect $S\sim L^{d_{\min}}$ with $d_{\min}>1$. Instead, the data point to a marginal deviation from linear scaling.

To identify the appropriate form, we define the quantity $s_p\equiv (S/L)^p$ and search for the value of $p$ that produces the cleanest large-$L$ collapse against $\ln L$. Figure~\ref{s4_gap}(a) shows that $p=4$ gives an asymptotically linear collapse, which is equivalent to
\begin{align}
    S\sim L (\ln L)^{1/4}.
    \label{chem}
\end{align}
This is consistent with the Ding-Wirth upper bound for the chemical distance in two-dimensional GFF level-set percolation\cite{ding2020}. We note that the sRW observable---the maximal breadth-first-search distance from the seed on the torus---differs in detail from the GFF crossing distance between separated boundaries; nevertheless, the agreement in functional form is suggestive. The data indicate that the sRW cluster contains spanning paths whose length grows only as $L(\ln L)^{1/4}$---i.e., linearly in $L$ up to a slowly divergent logarithmic factor---despite the highly perforated geometry of the trace.

The next important issue is whether Eq.~\eqref{chem} should be multiplied by an additional double-logarithmic correction. We therefore consider the more general possibility
\begin{equation}
    S/L\sim (\ln L)^{1/4}(\ln\ln L)^{m},
\end{equation}
where $m$ is an exponent that quantifyes the strength of the double-logarithmic correction.
To probe this possibility, we define the gap function with $s_4\equiv (S/L)^4$,
\begin{align}
    \Delta_{s_4}(L)\equiv s_4(2L)-s_4(L),
    \label{gapfun}
\end{align}
and study its asymptotic evolution.
According to the scaling form, $\Delta_{s_4}(L)$ is expected to be asymptotically convergent for $m=0$, to drift upward for $m>0$, and to drift downward for $m<0$.
As plotted against $\ln L$ in Fig.~\ref{s4_gap}(b), the gap function shows no clear evidence for asymptotic convergence or divergence over the accessible system sizes ($L\leq 2^{16}$).
This observation argues against the existence of a $(\ln\ln L)^m$ correction with $m\ge0$.
Instead, the data exhibit a sustained decreasing trend for large $L$ (Fig.~\ref{s4_gap}(b)), which is not clearly discernible within statistical errors.
We note, however, that $\ln\ln L$ varies very slowly over this limited range, so the current data are not sufficient to exclude a small negative value of $m$. Consequently, within our numerical precision, the data are consistent with the simpler scaling form given by Eq.~\eqref{chem}, and we find no detectable $(\ln\ln L)^m$ correction.

For a quantitative fit, we use
\begin{align}
    S(L)=L{(c+\ln L)^{k}}(a+bL^{-\omega}),
    \label{c_scaling}
\end{align}
where $\omega$ describes a subleading correction. The results in Table \ref{chem_fit} are consistent with $k=1/4$. When the correction term is omitted, the fitted exponent drifts toward $0.25$ as $L_{\min}$ increases. When $k=1/4$ and $\omega=1$ are fixed, the remaining amplitudes are stable within error bars. The overall picture is therefore plausible: the shortest spanning path scales linearly in $L$ with a weak logarithmic enhancement $(\ln L)^{1/4}$.

\begin{table}[t]
    \centering
    \caption{Fits of the chemical distance to Eq.~\eqref{c_scaling}.}
    \label{chem_fit}
    \begin{ruledtabular}
        \begin{tabular}{c c S[table-format=1.3(2)] S[table-format=-1.1(2)] S[table-format=1.4(2)]
                S[table-format=-1.1(2)] c}
            {$L_{\min}$} & $\chi^2/\mathrm{DF}$ & {$k$}     & {$c$}     & {$a$}      & {$b$}    & {$w$} \\
            \midrule
            512          & \chidf{7.9}{5}       & 0.249(4)  & 0.17(11)  & 1.086(12)  & {--}     & --    \\
            1024         & \chidf{2.6}{4}       & 0.238(6)  & -0.21(19) & 1.124(20)  & {--}     & --    \\
            2048         & \chidf{0.3}{3}       & 0.225(10) & -0.7(4)   & 1.17(4)    & {--}     & --    \\
            4096         & \chidf{0.1}{2}       & 0.231(20) & -0.5(8)   & 1.15(8)    & {--}     & --    \\
            \midrule
            512          & \chidf{7.9}{5}       & {1/4}     & 0.20(3)   & 1.0823(8)  & 0.04(17) & 1     \\
            1024         & \chidf{4.0}{4}       & {1/4}     & 0.26(5)   & 1.0805(12) & -0.8(5)  & 1     \\
            2048         & \chidf{0.3}{3}       & {1/4}     & 0.38(8)   & 1.077(3)   & -2.9(12) & 1     \\
            4096         & \chidf{0.3}{2}       & {1/4}     & 0.39(15)  & 1.077(4)   & -3(4)    & 1     \\
        \end{tabular}
    \end{ruledtabular}
\end{table}

\section{Conclusions}
We have studied three geometric observables of the cluster generated by a 2D sRW of $L^2$ steps on the periodic square lattice. First, the mass follows the Dvoretzky-Erd\H{o}s law $M\simeq (\pi/2)L^2/\ln L$, while the dominant visible finite-size correction under PBCs is of order $(\ln L)^{-2}$. Second, the hull perimeter obeys a clean power law with $d_{\rm hull}=1.333\,29(14)$, in agreement with the exact Brownian-frontier value $4/3$ and hence with $\mathrm{SLE}_{8/3}$\cite{lawler2001,lawler2001b,lawler2001c,lawler2002,werner2004}. Third, the spanning chemical distance is consistent with the near-linear form $S\sim L(\ln L)^{1/4}$, showing that the cluster contains highly efficient spanning paths despite its perforated geometry.

The finite-size analysis also reveals a consistent role of periodic boundary conditions. For the mass, PBCs leave the leading coefficient $\pi/2$ intact, but the dominant visible correction is $(\ln L)^{-2}$ rather than the $(\ln L)^{-1}$ structure suggested by the infinite-plane analysis\cite{csaki2024}. For the hull, PBCs affect the observable $L^{-1}$ correction, while the leading exponent remains consistent with $4/3$ within our precision. These results suggest that PBCs primarily modify the approach to scaling rather than the leading asymptotic behavior. It is therefore natural to expect that the same is true for the chemical distance.

This point is particularly relevant to the Ding-Wirth upper bound $L(\ln L)^{1/4}$ for the chemical distance in two-dimensional GFF level-set percolation\cite{ding2020}. Our gap-function analysis shows no detectable evidence for an additional $(\ln\ln L)^m$ factor over the accessible range of system sizes, and the fits with $k=1/4$ remain stable once subleading corrections are included. Thus, the present data support the conjecture that the bound is sharp for the sRW cluster as well. This conjecture is further motivated by the structural connection between random-walk occupation fields and the GFF through isomorphism theorems\cite{dynkin1983,sznitman2012,lejan2011,lupu2016}, which suggests that the two models share closely related geometric scaling properties.

The chemical-distance result can be viewed from another angle by unwrapping the periodic trajectory onto the infinite plane. In that representation, periodic shortcuts are removed, so the corresponding infinite-plane mass, hull perimeter, and spanning shortest-path scale are naturally expected to be no smaller than their periodic counterparts. In this sense, the torus estimate of the chemical distance is conservative: the fact that it already exhibits the form $L(\ln L)^{1/4}$ lends further support to the view that this is the correct asymptotic order, rather than a finite-size underestimate masking an additional $(\ln\ln L)^m$ factor. While this argument is heuristic and not a proof, it is fully consistent with the sharpness conjecture above.

Taken together, these results provide a coherent fractal picture of the 2D sRW cluster: marginal mass scaling, a Brownian-frontier exterior, and nearly optimal internal connectivity. It would be interesting to extend this analysis to other lattice geometries and boundary conditions, and to explore related questions for loop-erased random walks and other 2D conformally invariant random geometries.

\begin{acknowledgments}
    We are grateful to Xin Sun, Xinyi Li, Jian Ding and Shuo Wei for useful discussions. This work receives extensive guidance from Youjin Deng. P.H. acknowledges the support by the National Natural Science Foundation of China (NSFC) under Grant No. 12204173 and No. 12275263, as well as the Innovation Program for Quantum Science and Technology (under Grant No. 2021ZD0301900).
\end{acknowledgments}

\bibliographystyle{apsrev4-2}
\bibliography{ref}

\end{document}